\documentclass[a4paper]{article}
\usepackage{float}
\restylefloat{figure}
\restylefloat{table}

\usepackage{amsmath, amssymb, eqparbox}
\usepackage[margin=20mm]{geometry}
\usepackage{algorithm}
\usepackage{algpseudocode}
\usepackage{graphicx}
\usepackage{multirow}
\usepackage{url}
\usepackage{pdfpages}
\usepackage{hyperref}
\usepackage{caption}
\usepackage{float}

\usepackage[round]{natbib}
\providecommand{\keywords}[1]
{
  \small	
  \textbf{\textit{Keywords---}} #1
}

\DeclareMathOperator*{\argmin}{arg\,min}

\floatstyle{plaintop}
\restylefloat{table}

\title{Collaborative causal inference on distributed data}

\author{Yuji Kawamata~\thanks{Center for Artificial Intelligence Research, University of Tsukuba, Tsukuba, Japan, \texttt{yjkawamata@gmail.com}} \and 
Ryoki Motai~\thanks{Graduate School of Science and Technology, University of Tsukuba, Tsukuba, Japan, \texttt{motai.ryoki.tg@alumni.tsukuba.ac.jp}} \and
Yukihiko Okada~\thanks{Center for Artificial Intelligence Research, University of Tsukuba, Tsukuba, Japan, \texttt{okayu@sk.tsukuba.ac.jp}} \and
Akira Imakura~\thanks{Center for Artificial Intelligence Research, University of Tsukuba, Tsukuba, Japan, \texttt{imakura@cs.tsukuba.ac.jp}} \and
Tetsuya Sakurai~\thanks{Center for Artificial Intelligence Research, University of Tsukuba, Tsukuba, Japan, \texttt{sakurai@cs.tsukuba.ac.jp}}
}

\date{\today}

\begin{document}
\pagebreak \setcounter{page}{1}
\pagenumbering{arabic} 

\maketitle

\begin{abstract}
In recent years, the development of technologies for causal inference with privacy preservation of distributed data has gained considerable attention.
Many existing methods for distributed data focus on resolving the lack of subjects (samples) and can only reduce random errors in estimating treatment effects.
In this study, we propose a data collaboration quasi-experiment (DC-QE) that resolves the lack of both subjects and covariates, reducing random errors and biases in the estimation.
Our method involves constructing dimensionality-reduced intermediate representations from private data from local parties, sharing intermediate representations instead of private data for privacy preservation, estimating propensity scores from the shared intermediate representations, and finally, estimating the treatment effects from propensity scores.
Through numerical experiments on both artificial and real-world data, we confirm that our method leads to better estimation results than individual analyses.
While dimensionality reduction loses some information in the private data and causes performance degradation, we observe that sharing intermediate representations with many parties to resolve the lack of subjects and covariates sufficiently improves performance to overcome the degradation caused by dimensionality reduction. 
Although external validity is not necessarily guaranteed, our results suggest that DC-QE is a promising method.
With the widespread use of our method, intermediate representations can be published as open data to help researchers find causalities and accumulate a knowledge base.
\end{abstract}


\section{Introduction }
\label{sec:introduction}

Quasi-experiments for causal inference are often used when experiments, such as randomized control trials, are difficult, particularly in the social and medical sciences.
Most quasi-experimental studies analyze their data individually to estimate treatment effects using statistical causal inference methods.
However, data that are centralized and analyzed in one place 
might provide a more reliable estimate of the treatment effect than that from distributed or individual analyses.
However, sharing data is difficult owing to privacy concerns.
Conversely, in the field of machine learning, federated learning has recently been proposed for analyzing distributed data without sharing individual private data and has gained significant attention \citep{konevcny2016federated,mcmahan2017communication}.
In addition, a new approach to federated learning systems has been proposed for sharing dimensionality-reduced intermediate representations instead of individual private data.
This approach is known as collaborative data analysis \citep{imakura2020data,Bogdanova2020federated,imakura2021accuracy,imakura2021interpretable}.
In this study, we propose a new causal inference framework based on collaborative data analysis.
Our method enables privacy-preserving statistical causal inference using propensity score methods \citep{rosenbaum1983central}.

\begin{figure}[tb]
  \centering
  \includegraphics[width=\linewidth]{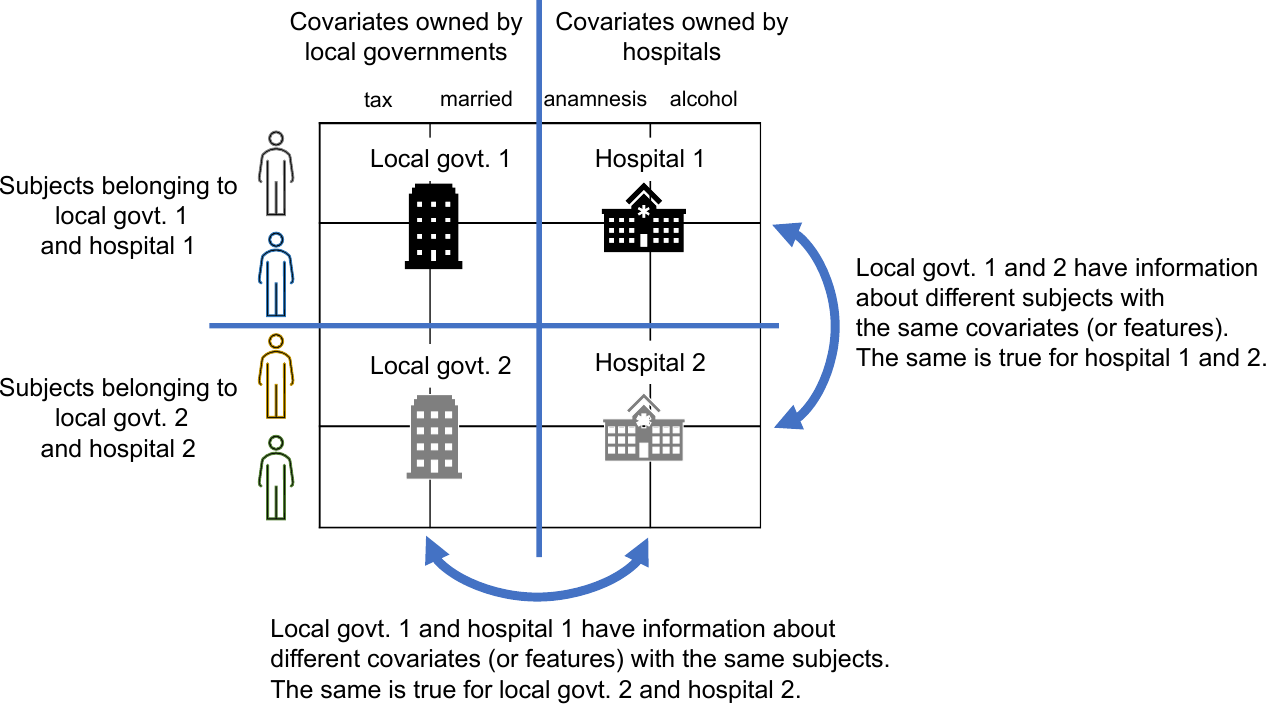}
  \caption{Example situation for distributed data.}
  \label{fig:setting}
\end{figure}

In the social sciences, a motivating example is the estimation of the effect of job training support on residents' incomes.
Fig. \ref{fig:setting} presents the data used in this example.
Each local government has statistics on follow-up support provided to residents.
A single local government may not obtain reliable estimation results owing to the limited number of subjects (or samples) and covariates (or features).
The effect of support may be estimated with less random error if data from  residents of each local government could be centralized to increase the sample size,
Moreover, if hospital-held medical information on residents could be centralized to increase the number of covariates, the treatment effect might be estimated with less bias.
A similar discussion may be had in the medical sciences, for example, when estimating pharmacological effects using distributed data.
However, due to privacy issues, sharing residents' data is difficult.
This difficulty might be solved by developing a method for estimating treatment effects from distributed data while maintaining privacy.

The development of federated learning provides the potential to develop causal inference methods with privacy preservation.
However, most studies related to federated learning have focused on regression or classification, with few studies focusing on statistical causal inference.
Moreover, the methods proposed in subsequent investigations have been based on specific parametric models and estimated treatment effects from data where subjects were distributed.
To the best of our knowledge, no method has been proposed to address statistical causal inference on private data where not only subjects but also covariates were distributed.
Developing statistical causal inference methods for such distributed data would reduce random error and bias in estimates.

Hence, we propose a collaborative causal inference framework based on the collaborative data analysis by \cite{imakura2020data}.
Our method can address estimation issues such as those mentioned as motivating examples.
We demonstrated through numerical experiments that our method, which can address data where covariates are distributed, leads to a remarkable improvement in performance owing to bias reduction.
Existing methods cannot utilize such distributed data.

The remainder of this paper is organized as follows.
In Section \ref{sec:relatedwork}, we present related work.
We describe the basics of causal inference and distributed data in Section \ref{sec:preliminaries}.
We describe our method in Section \ref{sec:method}.
In Section \ref{sec:experiments}, we report the numerical experiments and discuss the results in Section \ref{sec:discussion}.
Finally, we conclude in Section \ref{sec:conclusion}.


\section{Related work}
\label{sec:relatedwork}

Statistical causal inference, based on \cite{hernan2019second}, is the statistical prediction of outcomes (outputs) if treatments (inputs) in the world were different.
Most studies on statistical causal inference assume a situation where data can be dealt with as a single dataset
(e.g., \cite{imbens2015causal,glymour2016causal}).
Analyzing datasets from multiple studies can be useful for increasing the statistical power to detect treatment effects.
However, multiple datasets may not be shared because of practical constraints, such as privacy considerations.
The most widely used method to address these issues is meta-analysis, which provides a weighted average of individual studies (e.g., \cite{hartung2011statistical}).
However, meta-analyses have limitations, such as incomplete data reporting and publication bias \citep{gurevitch1999statistical}.
One solution to these limitations is to centralize all datasets \citep{huedo2012effectiveness}; however, this is not always possible.

Recent studies have developed privacy-preserving methods to pool summary-level information across multiple studies instead of using the original datasets.
These methods operate with specific parametric models, such as linear \citep{toh2018combining,toh2020privacy}, logistic \citep{duan2020learning}, Poisson \citep{shu2019privacy}, and Cox \citep{shu2020inverse,shu2021variance}.

Statistical causal inference has recently begun to be addressed in federated learning literature.
Pioneering works addressing statistical causal inference in the literature are, for example, \cite{xiong2021federated}, \cite{vo2022bayesian}, \cite{han2021federated} and \cite{han2022privacy}.
These studies have proposed privacy-preserving methods for statistical causal inference referred to as federated causal inferences.
\cite{xiong2021federated} considered aggregating individually estimated gradients for estimating propensity scores and treatment effects.
\cite{vo2022bayesian} proposed a potential outcome model based on Gaussian processes that is updated iteratively by communication.
\cite{han2021federated,han2022privacy} proposed a method to estimate average treatment effects for specific target populations by optimally weighting datasets according to their heterogeneity.
These methods can address horizontally partitioned data, where subjects are distributed.
In such cases, by estimating with more subjects, the methods can reduce random errors.
However, vertically partitioned data, where covariates are distributed, are also important for practical purposes, but cannot be addressed by these works.
Alternatively, federated learning methods have been proposed that can address regression or classification tasks in both types of partitioned data \citep{konevcny2016federated,mcmahan2017communication,Bogdanova2020federated,imakura2021interpretable,cheng2021secureboost,liu2022fedbcd}.

The most innovative aspect of our method compared to existing methods is that it can perform statistical causal inference for vertically partitioned data.
Estimating with more covariates, our method can reduce the omitted variable bias \citep{cinelli2020making}.
Table \ref{tab:relatedworks} summarizes the relationship between the existing federated learning studies and this study.
We also note that our method can estimate horizontally partitioned data.

\begin{table}[tb]
\centering
    \caption{Existing federated learning literature and this study}
    \begin{tabular}{lll}
    \hline\hline
    Task \textbackslash~ Partition & Horizontal (Subjects) & Vertical (Covariates) \\ \hline\hline
    \begin{tabular}[c]{@{}l@{}}Classification or \\ regression\end{tabular} & \begin{tabular}[c]{@{}l@{}}\cite{konevcny2016federated}\\ \cite{mcmahan2017communication}\\ \cite{Bogdanova2020federated}\end{tabular} & \begin{tabular}[c]{@{}l@{}}\cite{imakura2021interpretable}\\ \cite{cheng2021secureboost}\\ \cite{liu2022fedbcd}\end{tabular} \\ \hline
    Causal inference & \begin{tabular}[c]{@{}l@{}}\cite{vo2022bayesian}\\ \cite{xiong2021federated}\\ \cite{han2021federated}\\ \cite{han2022privacy}\\ \textbf{This study}\end{tabular} & \textbf{This study} \\ 
    \hline\hline
    \end{tabular}
\label{tab:relatedworks}
\end{table}


\section{Preliminaries}
\label{sec:preliminaries}

In this section, we present preliminary information that are needed for our method.
We begin with a description of treatment effects and propensity scores in Section \ref{sec:preliminaries_teps}.
Next, we present the data distribution settings in Section \ref{sec:preliminaries_dist}.

\subsection{Treatment effect and propensity score}
\label{sec:preliminaries_teps}
For simplicity, we consider here a situation where data are not distributed.
Let $m$ and $n$ be the numbers of covariates and subjects, respectively.
Additionally, let $X = [ \boldsymbol{x}_1, \boldsymbol{x}_2, \dots, \boldsymbol{x}_n ]^T \in \mathbb{R}^{n \times m}$,
$Z = [ z_1, z_2, \dots , z_n ]^T \in \{ 0,1 \}^n$ and
$Y = [ y_1, y_2, \dots , y_n ]^T \in \mathbb{R}^{n} $ be the covariates, treatments, and outcomes, respectively.

Under the Neyman--Rubin causal model (e.g., \cite{imbens2015causal}), assuming the stable unit treatment value assumption, let $y_i (1)$ and $y_i (0)$ be the outcomes that subject $i$ would have if they were treated ($z_i=1$) or not (controlled; $z_i=0$), respectively.
We consider two types of treatment effects: the average treatment effect (ATE) and the average treatment effect on the treated (ATT) as
\begin{align}
    \tau_\text{ATE} &= E[ y_i (1) - y_i (0) ], \\
    \tau_\text{ATT} &= E[ y_i (1) - y_i (0) | z_i=1].
\end{align}
We cannot observe the treatment effects because observing both $y_i (1)$ and $y_i (0)$ for subject $i$ is impossible.
This is the fundamental problem in causal inference \citep{holland1986statistics}.

However, under the strongly ignorable treatment assignment assumption, \cite{rosenbaum1983central} has shown that estimating treatment effects using the propensity score is possible.
\begin{equation}
    e_i = Pr(z_i=1 | \boldsymbol{x}_i ).
\end{equation}
Treatment and covariates are conditionally independent given the propensity score; that is,
\begin{equation}
    \boldsymbol{x}_i \perp\!\!\!\perp z_i | e_i,
\end{equation}
where $\perp\!\!\!\perp$ denotes that the variables are independent.
Several methods have been proposed to estimate treatment effects taking advantage of the fact that covariates are balanced by propensity scores.
In most cases, because the true propensity score $e_i$ is unknown, the estimated propensity score $\hat{e}_i$ is used instead to estimate treatment effects.

This study focuses on propensity score matching (PSM) and inverse probability weighting (IPW), which are representative methods for estimating treatment effects \citep{austin2011introduction}.

\subsubsection*{Propensity score matching (PSM)}
PSM pairs each participant in one group with a participant in the other group based on their propensity scores.
Although there are various methods for PSM \citep{ankarali2009comparison}, we consider the simplest method, one-to-one nearest-neighbor matching with replacement.
\footnote{This study does not consider the caliper, which is the acceptable upper limit of difference in propensity scores between matched subjects.}
The procedure for estimating treatment effects using PSM in this study is as follows.
We denote the treatment and control subjects by $\mathbb{N}_{T} = \{ i | z_i=1 \}$ and $\mathbb{N}_{C} = \{ i | z_i=0 \}$, respectively.
The comparison of each subject is determined by
\begin{equation}
    pair(i) = \begin{cases}
                    \argmin_{j \in \mathbb{N}_{C}} |\hat{e}_i-\hat{e}_j| & (\text{if}~ i \in \mathbb{N}_{T}) \\
                    \text{or} \\
                    \argmin_{j \in \mathbb{N}_{T}} |\hat{e}_i-\hat{e}_j| & (\text{if}~ i \in \mathbb{N}_{C})
              \end{cases}.
\end{equation}
Then, the estimates of ATE and ATT in PSM are
\begin{align}
    \hat{\tau}_\text{ATE}^\text{PSM} &= \frac{1}{n} \left( \sum_{i \in \mathbb{N}_{T}} \left( y_i - y_{pair(i)} \right) + \sum_{i \in \mathbb{N}_{C}} \left( y_{pair(i)} - y_i \right) \right), \\
    \hat{\tau}_\text{ATT}^\text{PSM} &= \frac{1}{|\mathbb{N}_{T}|} \sum_{i \in \mathbb{N}_{T}} \left( y_i - y_{pair(i)} \right).
\end{align}

\subsubsection*{Inverse probability weighting (IPW)}
IPW estimates treatment effects by weighting each outcome as the inverse of the propensity score.
For ATE, the weights are calculated as $\frac{1}{\text{propensity score}}$ for the treatment group and $\frac{1}{1-\text{propensity score}}$ for the control group.
For ATT, the weights are calculated as $1$ for the treatment group and $\frac{\text{propensity score}}{1-\text{propensity score}}$ for the control group.
Formally, the estimates of ATE and ATT in IPW are
\begin{align}
    \hat{\tau}_\text{ATE}^\text{IPW} &= \frac{ \sum_{i=1}^n \frac{z_i}{\hat{e}_i}y_i }{ \sum_{i=1}^n \frac{z_i}{\hat{e}_i} } -
    \frac{ \sum_{i=1}^n \frac{1-z_i}{1-\hat{e}_i}y_i }{ \sum_{i=1}^n \frac{1-z_i}{1-\hat{e}_i} }, \\
    \hat{\tau}_\text{ATT}^\text{IPW} &= \frac{ \sum_{i=1}^n z_i y_i }{ \sum_{i=1}^n z_i } - 
    \frac{ \sum_{i=1}^n \frac{(1-z_i)\hat{e}_i}{1-\hat{e}_i}y_i }{ \sum_{i=1}^n \frac{(1-z_i)\hat{e}_i}{1-\hat{e}_i} }.
\end{align}

\subsection{Distributed data setting}
\label{sec:preliminaries_dist}

Two types of data partitioning are often considered for distributed data analysis: horizontal and vertical \citep{imakura2021interpretable}.
Horizontal partitioning implies that different subjects with the same covariates are owned by multiple parties, whereas vertical partitioning implies that different covariates with the same subjects are owned by multiple parties.
Most studies on causal inference using distributed data consider only horizontal settings.
However, our method can be applied to more complicated situations where data are distributed horizontally and vertically, as shown in Fig. \ref{fig:setting}.

We consider that $n$ subjects are partitioned into $c$ institutions and the $m$ covariates are partitioned into $d$ parties, as follows:
\begin{align}
    X &= \begin{bmatrix}
            X_{1,1} & X_{1,2} & \cdots & X_{1,d} \\
            X_{2,1} & X_{2,2} & \cdots & X_{2,d}\\
            \vdots & \vdots & \ddots & \vdots \\
            X_{c,1} & X_{c,2} & \cdots & X_{c,d}
        \end{bmatrix},\\
    Z &= \begin{bmatrix}
            Z_1 \\ Z_2 \\ \vdots \\ Z_c
        \end{bmatrix},\\
    Y &= \begin{bmatrix}
            Y_1 \\ Y_2 \\ \vdots \\ Y_c
        \end{bmatrix}.
\end{align}
Then, the $(k,l)$th party has a partial dataset and corresponding treatments and outcomes, which are
$X_{k,l} \in \mathbb{R}^{n_k \times m_l}, Z_k \in \{ 0,1 \}^{n_k}, Y_k \in \mathbb{R}^{n_k}$,
where $n_k$ is the number of subjects for each party in the $k$th row ($n=\sum_{k=1}^c n_k$)
and $m_l$ is the number of covariates for each party in the $l$th column ($m=\sum_{l=1}^d m_l$).

Individual analysis of a dataset from a local party may not accurately estimate treatment effects
because a lack of covariates causes bias, and insufficient subjects increase random error.
If the datasets from multiple parties can be centralized and analyzed as a single dataset, that is, centralized analysis, 
accurate estimates can be expected.
However, individual data for centralization are difficult to share
owing to privacy and confidentiality concerns.

\section{Method}
\label{sec:method}

In this section, we first describe our method in Section \ref{sec:method_dcqe}.
Next, we discuss the privacy and confidentiality preservation schemes on our method in Section \ref{sec:method_privacy}.
Then, we present the advantages and disadvantages of our method in Section \ref{sec:method_ad_disad}.

\subsection{Data collaboration quasi-experiment}
\label{sec:method_dcqe}

Our method is based on collaborative data analysis proposed by \cite{imakura2020data} and enables causal inference with privacy preservation of distributed data.
In keeping with \cite{imakura2020data}'s term ``data collaboration,'' we refer to our method as a data collaboration quasi-experiment (DC-QE).
Here, we briefly explain the algorithm of collaborative data analysis in Algorithm \ref{alg-cd} before we explain our method, DC-QE.
Collaborative data analysis is a method originally proposed for predictive tasks.
Hence, the procedure of collaborative data analysis consists of training and prediction phases.
In the training phase, users (parties) work collaboratively to create a prediction model.
Then, in the prediction phase, users obtain predictions for their test dataset using the prediction model.
Refer to \cite{imakura2020data} for a detailed explanation of collaborative data analysis.

As in \cite{imakura2020data}, DC-QE operates in two roles: \textit{user} and \textit{analyst} roles.
Users have a private dataset $X_{k,l}$, treatments $Z_k$ and outcomes $Y_k$, which must be analyzed without sharing $X_{k,l}$.
DC-QE is conducted in two stages: construction of collaborative representations and estimation of treatment effects.
In the first stage, each user individually constructs intermediate representations and shares them with the analyst instead of the private dataset.
In the second stage, the analyst estimates the treatment effects.

\begin{algorithm}[tb]
    \caption{Collaborative data analysis}
    \label{alg-cd}
    \begin{algorithmic}[1]
        \Statex \textbf{Input: feature dataset $X$, outcomes $Y$}.
        \Statex \textbf{Output: prediction results $Y^\text{test}$}.
        \vspace{-.2\baselineskip}
        
        \Statex \hrulefill
        \vspace{-.3\baselineskip}
        \Statex \textit{user-side} $(k,l)$ (Training phase)
        \vspace{-.5\baselineskip}
        \Statex \hrulefill
        \vspace{-.1\baselineskip}
        \State Generate anchor dataset $X_{k,l}^\text{anc}$ and share it with all users.
        \State Set $X_{:,l}^\text{anc}$.
        \State Generate $f_{k,l}$.
        \State Compute $\widetilde{X}_{k,l} = f_{k,l}(X_{k,l})$.
        \State Compute $\widetilde{X}_{k,l}^\text{anc} = f_{k,l}(X_{:,l}^\text{anc})$.
        \State Share $\widetilde{X}_{k,l}$, $\widetilde{X}_{k,l}^\text{anc}$ and $Y_k$ to the analyst.
        \vspace{-.3\baselineskip}
        
        \Statex \hrulefill
        \vspace{-.3\baselineskip}
        \Statex \textit{analyst-side} (Training phase)
        \vspace{-.5\baselineskip}
        \Statex \hrulefill
        \vspace{-.1\baselineskip}
        \State Get $\widetilde{X}_{k,l}$, $\widetilde{X}_{k,l}^\text{anc}$ and $Y_k$ for all $k$ and $l$.
        \State Set $\widetilde{X}_{k}$ and $\widetilde{X}_{k}^\text{anc}$.
        \State Compute $g_k$ from $\widetilde{X}_{k}^\text{anc}$ for all $k$ such that $g_k(\widetilde{X}_{k}^\text{anc}) \approx g_{k'}(\widetilde{X}_{k'}^\text{anc}) ~ (k \neq k')$.
        \State Compute $\check{X}_k = g_k(\widetilde{X}_{k})$ for all $k$.
        \State Set $\check{X}$ and $Y$.
        \State Analyze $\check{X}$ and get prediction model $h$ such that $h(\check{X}) \approx Y$.
        \vspace{-.3\baselineskip}

        \Statex \hrulefill
        \vspace{-.3\baselineskip}
        \Statex \textit{user-side} $(k,l)$ (Prediction phase)
        \vspace{-.5\baselineskip}
        \Statex \hrulefill
        \vspace{-.1\baselineskip}
        \State Compute $\widetilde{X}_{k,l}^\text{test} = f_{k,l}(X_{k,l}^\text{test})$.
        \State Share $\widetilde{X}_{k,l}^\text{test}$ to the analyst.
        \vspace{-.3\baselineskip}
        
        \Statex \hrulefill
        \vspace{-.3\baselineskip}
        \Statex \textit{analyst-side} (Prediction phase)
        \vspace{-.5\baselineskip}
        \Statex \hrulefill
        \vspace{-.1\baselineskip}
        \State Get $\widetilde{X}_{k,l}^\text{test}$ for all $k$ and $l$.
        \State Set $\widetilde{X}_{k}^\text{test}$.
        \State Compute $Y_k^\text{test} = h(g_k(\widetilde{X}_{k}^\text{test}))$ for all $k$.
        \State Return $Y_k^\text{test}$ to users.
        \vspace{-.3\baselineskip}
        
        \Statex \hrulefill
        \vspace{-.3\baselineskip}
        \Statex \textit{user-side} $(k,l)$ (Prediction phase)
        \vspace{-.5\baselineskip}
        \Statex \hrulefill
        \vspace{-.1\baselineskip}
        \State Get $Y_k^\text{test}$.
        
        \end{algorithmic}
\end{algorithm}

\begin{figure}[tb]
  \centering
  \includegraphics[width=\linewidth]{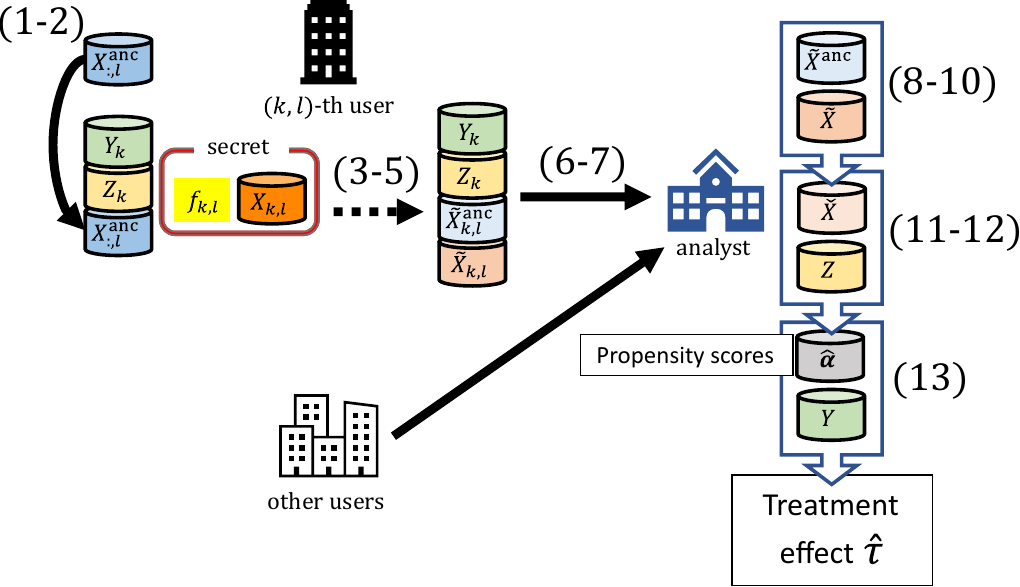}
  \caption{Overall illustration of the proposed method.}
  \label{fig:method}
\end{figure}

The first and second stages are described in detail in Sections \ref{sec:method_dcqe_const} and \ref{sec:method_dcqe_est}, respectively. 
The pseudocode of DC-QE, described below, is given in Algorithm \ref{alg-ps-cd}.
Fig. \ref{fig:method} shows an overall illustration of DC-QE, whose numbers correspond to the line numbers in Algorithm \ref{alg-ps-cd}.

\begin{algorithm}[tb]
    \caption{Data collaboration quasi-experiment (DC-QE)}
    \label{alg-ps-cd}
    \begin{algorithmic}[1]
        \Statex \textbf{Input: covariate dataset $X$, treatments $Z$, outcomes $Y$}.
        \Statex \textbf{Output: estimated treatment effect $\hat{\tau}$}.
        \vspace{-.3\baselineskip}
        
        \Statex \hrulefill
        \vspace{-.3\baselineskip}
        \Statex \textit{user-side} $(k,l)$
        \vspace{-.5\baselineskip}
        \Statex \hrulefill
        \vspace{-.1\baselineskip}
        \State Generate anchor dataset $X_{k,l}^\text{anc}$ and share it with all users.
        \State Set $X_{:,l}^\text{anc}$.
        \State Generate $f_{k,l}$.
        \State Compute $\widetilde{X}_{k,l} = f_{k,l}(X_{k,l})$. (\ref{eq:trans_raw_ir})
        \State Compute $\widetilde{X}_{k,l}^\text{anc} = f_{k,l}(X_{:,l}^\text{anc})$. (\ref{eq:trans_anc_ir})
        \State Share $\widetilde{X}_{k,l}$, $\widetilde{X}_{k,l}^\text{anc}$, $Z_k$ and $Y_k$ to the analyst.
        \vspace{-.3\baselineskip}
        
        \Statex \hrulefill
        \vspace{-.3\baselineskip}
        \Statex \textit{analyst-side}
        \vspace{-.5\baselineskip}
        \Statex \hrulefill
        \vspace{-.1\baselineskip}
        \State Get $\widetilde{X}_{k,l}$, $\widetilde{X}_{k,l}^\text{anc}$, $Z_k$ and $Y_k$ for all $k$ and $l$.
        \State Set $\widetilde{X}_{k}$ and $\widetilde{X}_{k}^\text{anc}$.
        \State Compute $g_k$ from $\widetilde{X}_{k}^\text{anc}$ for all $k$ such that $g_k(\widetilde{X}_{k}^\text{anc}) \approx g_{k'}(\widetilde{X}_{k'}^\text{anc}) ~ (k \neq k')$. (\ref{eq:g_cond})
        \State Compute $\check{X}_k = g_k(\widetilde{X}_{k})$ for all $k$. (\ref{eq:trans_ir_dc})
        \State Set $\check{X}$, $Z$ and $Y$.
        \State Estimate propensity scores $\hat{\boldsymbol{\alpha}}$ from $\check{X}$ and $Z$.
        \State Estimate treatment effect $\hat{\tau}$ from $\hat{\boldsymbol{\alpha}}$ and $Y$ using existing ways (PSM, IPW, and so on).

        \end{algorithmic}
\end{algorithm}

\subsubsection{Construction of collaborative representations}
\label{sec:method_dcqe_const}

In the first stage, construction of collaborative representations, we use the aggregation method proposed in \cite{imakura2020data}.
Collaborative representations are constructed from dimensionality-reduced intermediate representations.
We provide a brief description (refer to \cite{imakura2021interpretable} for details).
First, users generate and share an anchor dataset $X^\text{anc} \in \mathbb{R}^{r \times m}$, which is a shareable dataset consisting of public  or randomly constructed dummy data, where $r$ is the number of subjects in the anchor data (line 1 of Alg. \ref{alg-ps-cd}).
The anchor dataset is used to compute the transformation functions (described below as $g_k$) required to construct the collaborative representations.
Second, each user constructs intermediate representations as
\begin{align}
    \widetilde{X}_{k,l} &= f_{k,l}(X_{k,l}) \in \mathbb{R}^{n_k \times \widetilde{m}_{k,l}}, \label{eq:trans_raw_ir} \\
    \widetilde{X}_{k,l}^\text{anc} &= f_{k,l}(X_{:,l}^\text{anc}) \in \mathbb{R}^{r \times \widetilde{m}_{k,l}} \label{eq:trans_anc_ir}
\end{align}
where $f_{k,l}$ is a linear or nonlinear row-wise dimensionality reduction function, $X_{:,l}^\text{anc}$ is a partitioned anchor dataset with the same covariates as the user, and $\widetilde{m}_{k,l} ~ (< m_{k,l})$ is the reduced dimension (lines 2--5 of Alg. \ref{alg-ps-cd}).
For the function $f_{k,l}$, various dimensionality reduction methods can be used, including unsupervised and supervised learning methods \citep{pearson1901liii,fisher1936use,he2003locality}.
Moreover, $f_{k,l}$ is private and can differ from other users.
Third, users share their intermediate representations, instead of the original private dataset $X_{k,l}$, and treatments and outcomes to the analyst (line 6 of Alg. \ref{alg-ps-cd}).

Next, the analyst transforms the shared intermediate representations into an incorporable form called collaborative representations.
To do this, the analyst computes the transformation functions for all $k$ such that
\begin{equation}
    g_k(\widetilde{X}_{k}^\text{anc}) \approx g_{k'}(\widetilde{X}_{k'}^\text{anc}) \in \mathbb{R}^{r \times \check{m}} ~ (k \neq k'),
    \label{eq:g_cond}
\end{equation}
where $\widetilde{X}_k^\text{anc} = [\widetilde{X}_{k,1}^\text{anc}, \widetilde{X}_{k,2}^\text{anc}, \dots, \widetilde{X}_{k,d}^\text{anc}]$ and 
$\check{m}$ is the dimension of collaborative representations (lines 7--9 of Alg. \ref{alg-ps-cd}).
Assuming $g_k$ is a linear function, we have the collaborative representation for $k$ as
\begin{equation}
    \check{X}_k = g_k(\widetilde{X}_{k}) = \widetilde{X}_{k} G_k \in \mathbb{R}^{n_k \times \check{m}}, \label{eq:trans_ir_dc}
\end{equation}
where $\widetilde{X}_{k} = [ \widetilde{X}_{k,1}, \widetilde{X}_{k,2}, \dots, \widetilde{X}_{k,d} ]$ and $G_k$ is the matrix representation of $g_k$ (line 10 of Alg. \ref{alg-ps-cd}).
Let
\begin{equation}
    \left[ \widetilde{X}_1^\text{anc}, \widetilde{X}_2^\text{anc}, \dots, \widetilde{X}_c^\text{anc} \right] =
        \begin{bmatrix}
            U_1 & U_2
        \end{bmatrix}
        \begin{bmatrix}
                \Sigma_1 &  \\
                 & \Sigma_2
            \end{bmatrix}
        \begin{bmatrix}
                V_1^T  \\
                V_2^T
        \end{bmatrix}
        \approx U_1 \Sigma_1 V_1^T
\end{equation}
be a low-rank approximation based on singular value decomposition of the matrix combining $\widetilde{X}_k^\text{anc}$,
where $\Sigma_1 \in \mathbb{R}^{\check{m} \times \check{m}}$ is a diagonal matrix whose diagonal entries are the larger parts of the singular values 
and $U_1$ and $V_1$ are column orthogonal matrices whose columns are the corresponding left and right singular vectors, respectively.
Matrix $G_k$ is then computed as
\begin{equation}
    G_k = \left( \widetilde{X}_k^\text{anc} \right)^\dagger U_1,
\end{equation}
where $\dagger$ denotes the Moore--Penrose inverse.
Then, the collaborative representations are given by
\footnote{
\cite{imakura2021interpretable} uses the hat for collaborative representations as $\hat{X}$.
However, because the hat is used to denote an estimated value in statistics, we use the check to avoid confusion.
}
\begin{equation}
    \check{X} =
    \begin{bmatrix}
        \check{X}_1 \\ \check{X}_2 \\ \vdots \\ \check{X}_c
    \end{bmatrix}
    \in \mathbb{R}^{n \times \check{m}}.
\end{equation}

\subsubsection{Estimation of treatment effects}
\label{sec:method_dcqe_est}

In the second stage, the estimation of treatment effects,
the analyst estimates treatment effects using the estimated treatment probabilities from $\check{X},Z$ and $Y$ (line 11 of Alg. \ref{alg-ps-cd}).
We define the treatment probability conditioned on the collaborative representation as
\begin{equation}
    \alpha_i = Pr( z_i=1 | \boldsymbol{\check{x}}_i ),
\end{equation}
where $\boldsymbol{\check{x}}_i$ is subject $i$'s collaborative representation in $\check{X}$.
In this study, we also refer to $\alpha_i$ as a propensity score unless misunderstandings occur.
Using methods such as logistic regression, the analyst can estimate $\alpha_i$ from $\check{X}$ and $Z$, which are centralized in the first stage (line 12 of Alg. \ref{alg-ps-cd}).
Studies related to collaborative data analysis \citep{imakura2020data,Bogdanova2020federated,imakura2021interpretable}
have shown that models using collaborative representations can be comparable to those using original private datasets in classification tasks.
This result suggests that $\hat{\alpha}_i$ may be a good approximation of $\hat{e}_i$.

The use of $\hat{\alpha}_i$ instead of $\hat{e}_i$ enables estimations using various methods proposed in existing propensity score studies. 
This study focuses on PSM and IPW, as described in Section \ref{sec:preliminaries_teps}.
The analyst estimates treatment effects based on the following (line 13 of Alg. \ref{alg-ps-cd}).

\subsubsection*{PSM based on DC-QE}
In DC-QE with PSM (hereinafter referred to as DC-QE(PSM)), the comparison of each subject is determined by
\begin{equation}
    pair^\text{DC-QE}(i) = \begin{cases}
                    \argmin_{j \in \mathbb{N}_{C}} |\hat{\alpha}_i-\hat{\alpha}_j| & (\text{if}~ i \in \mathbb{N}_{T}) \\
                    \text{or} \\
                    \argmin_{j \in \mathbb{N}_{T}} |\hat{\alpha}_i-\hat{\alpha}_j| & (\text{if}~ i \in \mathbb{N}_{C})
                \end{cases}.
\end{equation}
Then, the estimates of ATE and ATT in DC-QE(PSM) are
\begin{align}
    \hat{\tau}_\text{ATE}^\text{DC-QE(PSM)} &= \frac{1}{n} \left( \sum_{i \in \mathbb{N}_{T}} \left( y_i - y_{pair^\text{DC-QE}(i)} \right) + \sum_{i \in \mathbb{N}_{C}} \left( y_{pair^\text{DC-QE}(i)} - y_i \right) \right),\\
    \hat{\tau}_\text{ATT}^\text{DC-QE(PSM)} &= \frac{1}{|\mathbb{N}_{T}|} \sum_{i \in \mathbb{N}_{T}} \left( y_i - y_{pair^\text{DC-QE}(i)} \right).
\end{align}

\subsubsection*{IPW based on DC-QE}
Similarly, the estimates of ATE and ATT in DC-QE with IPW (hereinafter referred to as DC-QE(IPW)) are
\begin{align}
    \hat{\tau}_\text{ATE}^\text{DC-QE(IPW)} &= \frac{ \sum_{i=1}^n \frac{z_i}{\hat{\alpha}_i}y_i }{ \sum_{i=1}^n \frac{z_i}{\hat{\alpha}_i} } -
    \frac{ \sum_{i=1}^n \frac{1-z_i}{1-\hat{\alpha}_i}y_i }{ \sum_{i=1}^n \frac{1-z_i}{1-\hat{\alpha}_i} },\\
    \hat{\tau}_\text{ATT}^\text{DC-QE(IPW)} &= \frac{ \sum_{i=1}^n z_i y_i }{ \sum_{i=1}^n z_i } - 
    \frac{ \sum_{i=1}^n \frac{(1-z_i)\hat{\alpha}_i}{1-\hat{\alpha}_i}y_i }{ \sum_{i=1}^n \frac{(1-z_i)\hat{\alpha}_i}{1-\hat{\alpha}_i} }.
\end{align}

\subsection{Discussion on privacy and confidentiality preservation schemes}
\label{sec:method_privacy}

We discuss how privacy and confidentiality are preserved by the proposed method in the same manner as in existing data collaboration analyses \citep{imakura2020data,imakura2021interpretable,imakura2021accuracy}.
The proposed method preserves private dataset $X_{k,l}$ for both users and analyst under certain assumptions.
We assume that users do not trust each other and want to protect their private dataset $X_{k,l}$ from honest-but-curious users and the analyst.
We also assume that the analyst does not collude with any user.

First, we discuss how $X_{k,l}$ is preserved for other users.
In the proposed method, each user shares only the local anchor dataset $X_{k,l}^\text{anc}$ with other users.
The local anchor dataset is constructed by users using methods such as data augmentation and does not contain $X_{k,l}$.
Hence, other users cannot extract $X_{k,l}$ directly.

Second, we discuss how $X_{k,l}$ is preserved for the analyst.
In the proposed method, each user shares only the intermediate representations $\widetilde{X}_{k,l}$ and $\widetilde{X}_{k,l}^\text{anc}$ with the analyst.
The analyst cannot obtain an approximate function for $f_{k,l}$ because the function $f_{k,l}$ can be approximated only by someone who has both the input $X_{k,l}$ and output $\widetilde{X}_{k,l}$.
Hence, the analyst cannot infer $X_{k,l}$ using the inverse function of the approximate function for $f_{k,l}$.

\subsection{Advantages and Disadvantages}
\label{sec:method_ad_disad}

Here, we discuss the advantages and disadvantages of the proposed method.
The advantages of the proposed method are summarized in the following three points.
First, the proposed method can address data where subjects and covariates are distributed, whereas existing methods can address only data where subjects are distributed, as mentioned in Section \ref{sec:relatedwork}.
In particular, addressing data where covariates are distributed enables us to reduce bias owing to lack of covariates.
Second, as shown in Algorithm \ref{alg-ps-cd}, the proposed method is based on the one-path algorithm, which does not require iteration steps for data communication.
Third, researchers can select methods in the estimation, such as DC-QE(PSM) and DC-QE(IPW) according to their needs.

The disadvantages are summarized in the following two points.
First, dimensionality reduction may lose some information in the original dataset.
While this can cause performance degradation, sharing intermediate representations with many parties to resolve the lack of subjects and covariates may sufficiently improve performance to overcome degradation.
Second, the proposed method cannot be used directly if the outcome or treatment variable contains private information.
However, combining the proposed method with other privacy preservation methods, such as encryption technology, might fix the issue.
This study does not focus on this issue, but the issue will be addressed in future work.


\section{Experiments}
\label{sec:experiments}

In this section, we evaluate the performance of the proposed method through two numerical experiments.
Experiment I aims to confirm the proof-of-concept in artificial data, and Experiment II aims to evaluate the performance in real-world data.
We describe the common settings for these experiments and the evaluation scheme in Section \ref{sec:exp_settings}.
Then, we present Experiments I and II results in Sections \ref{sec:exp_I} and \ref{sec:exp_II}, respectively.
\footnote{We performed these experiments using Python codes, which are available from the corresponding author by reasonable request.}

\subsection{Common settings and evaluation scheme}
\label{sec:exp_settings}

We compared the performances of DC-QE (Algorithm \ref{alg-ps-cd}), centralized analysis, individual analysis (CA and IA, respectively, in Table \ref{tab:exp1},\ref{tab:exp2} and Fig. \ref{fig:exp1ate},\ref{fig:exp2att}), federated IPW-MLE \citep{xiong2021federated} and FedCI \citep{vo2022bayesian}.
In centralized analysis, the analyst has the entire dataset $X$, treatments $Z$, and outcomes $Y$, and estimates their treatment effects.
Conversely, in individual analysis, the analyst has one dataset $X_{k,l}$, treatments $Z_{k,l}$ and outcomes $Y_{k,l}$, and estimates the treatment effects from only them.
Note that centralized analysis is considered an ideal case because the datasets $X_{k,l}$ cannot be shared in our target situation.
Federated IPW-MLE (FIM) and FedCI are the related methods to this study, described in Section \ref{sec:relatedwork}.
We did not consider the method proposed in \cite{han2021federated,han2022privacy} because it aims to estimate the target average treatment effect (TATE), which is different from the estimands (ATE and ATT) that is the focus of this study.
We aim mainly to demonstrate the extent to which the proposed method outperforms individual analysis by comparing its performance with that of individual analyses. 

Data distribution settings were common in both experiments.
The entire dataset $X$ was distributed into four parties: $c=d=2$ as
\begin{equation}
    X = \begin{bmatrix}
        X_{1,1} & X_{1,2} \\
        X_{2,1} & X_{2,2}
    \end{bmatrix}.
\end{equation}
In this setting, five types of collaboration can be considered:
\begin{itemize}
    \item Left-side collaboration between (1,1)th and (2,1)th parties (L-clb in tables and figures),
    \item Right-side collaboration between (1,2)th and (2,2)th parties (R-clb in tables and figures),
    \item Top-side collaboration between (1,1)th and (1,2)th parties (T-clb in tables and figures),
    \item Bottom-side collaboration between (2,1)th and (2,2)th parties,
    \item Whole collaboration between all parties (W-clb in tables and figures).
\end{itemize}
In the experiments, we assume a situation where parties belonging to the same collaboration estimate treatment effects according to Algorithm \ref{alg-ps-cd}.
We explain details of data distribution for each experiment in Sections \ref{sec:exp_I} and \ref{sec:exp_II}.

We set the following conditions for the experiments.
The anchor dataset $X^\text{anc}$ consisted of random numbers drawn from uniform distributions with the maximum and minimum values of the entire dataset as intervals.
This type of anchor data was used by \cite{imakura2020data}.
The number of subjects $r$ in the anchor data was equal to $n$.
Each party used principal component analysis (PCA) \citep{pearson1901liii} after standardization as a dimensionality reduction function.
PCA is one of the most widely used dimensionality reduction methods.
The analyst estimated propensity scores using logistic regression, which is often used in the literature on propensity scores.
The logistic regression model used in the proposed method consisted of a linear combination of collaborative representations and a constant term.
Similarly, the logistic regression model for central or individual analysis consisted of a linear combination of covariates and a constant term.

We considered the following for the related methods.
For FIM, we used linear regression with Gaussian error and logistic regression as outcome and propensity models, respectively; moreover, we assumed that both models were stable (refer to \cite{xiong2021federated} for details), and each party used only its own variables for its models.
For FedCI, we used default hyperparameters of the author's code\footnote{\url{https://github.com/vothanhvinh/FedCI}}, i.e., 2000 for the number of iterations and $10^{-3}$ for learning rate.
Moreover, we considered left or right-side collaboration for FIM and FedCI because those can address only horizontally partitioned data.

We evaluated the performance of each method in terms of three aspects: accuracy of estimated treatment effect, inconsistency of estimated propensity score, and covariate balance.
\subsubsection*{Accuracy of estimated treatment effect}
We estimated the treatment effects from DC-QE(PSM) and DC-QE(IPW) for the proposed method and from PSM and IPW for central and individual analyses.
We investigated the distributions of treatment effects with $B=1000$ bootstrap replicates for all analyses except FedCI.
Owing to computational costs, results for FedCI were calculated with $B=100$.
We considered the performance of the estimated treatment effect as a gap in the bootstrap estimates with respect to a benchmark value $\tau^\text{BM}$ as
\begin{equation}
    \text{Gap}(\hat{\boldsymbol{\tau}},\tau^\text{BM}) =
    \sqrt{\frac{1}{B} \sum_{b=1}^{B} \left( \hat{\tau}_b - \tau^\text{BM} \right)^2},
\end{equation}
where $\hat{\boldsymbol{\tau}}=[\hat{\tau}_1,\dots,\hat{\tau}_{B}]^T$ and $\hat{\tau}_b$ is the bootstrap estimate of the $b$th replicate.
As described later, we set appropriate values as benchmarks $\tau^\text{BM}$ in Experiments I and II.

\subsubsection*{Inconsistency of estimated propensity score}
To evaluate the accuracy of the estimated propensity scores used to estimate the treatment effects,
we used metrics, inconsistency with true and CA, defined as
\begin{align}
    \text{InconsistencywithTrue}(\hat{\boldsymbol{e}}, \boldsymbol{e}^\text{True}) &= 
    \sqrt{\frac{1}{n} \sum_{i=1}^{n} \left( \hat{e}_i - e_i^\text{True} \right)^2}, \\
    \text{InconsistencywithCA}(\hat{\boldsymbol{e}}, \hat{\boldsymbol{e}}^\text{CA}) &= 
    \sqrt{\frac{1}{n} \sum_{i=1}^{n} \left( \hat{e}_i - \hat{e}_i^\text{CA} \right)^2},
\end{align}
where $\hat{\boldsymbol{e}} = [\hat{e}_1,\dots,\hat{e}_n]^T$ are the estimated propensity scores, 
$\boldsymbol{e}^\text{True} = [e_1^\text{True},\dots,e_n^\text{True}]^T$ are the true propensity scores, and
$\hat{\boldsymbol{e}}^\text{CA} = [\hat{e}_1^\text{CA},\dots,\hat{e}_n^\text{CA}]^T$ are the propensity scores estimated in the central analysis.
Note that $\hat{e}_i$ was replaced by $\hat{\alpha}_i$ in the proposed method.
The inconsistency with true was not calculated in Experiment II because true propensity scores were unknown in real-world data.
For FedCI, the inconsistencies with true and CA were not available because it does not calculate the propensity score.

\subsubsection*{Covariate balance}
Covariate balance is often evaluated using standardized mean differences \citep{austin2015moving}
\begin{equation}
    d^j = \frac{\bar{x}_{T}^j-\bar{x}_{C}^j}{\sqrt{ \frac{ s_{T}^j + s_{C}^j}{2} }}, \label{eq:smd}
\end{equation}
where $\bar{x}_{T}^j$ and $s_{T}^j$ are the mean and variance of covariate $j$ for the treatment group, respectively,
and $\bar{x}_{C}^j$ and $s_{C}^j$ are those of the control group.
For PSM or DC-QE(PSM), we estimated the standardized mean difference from (\ref{eq:smd}).
We can estimate those for IPW and DC-QE(IPW) by replacing the means and variances in (\ref{eq:smd}) with the weighted mean and variance.
The weighted mean and variance are defined as follows:
\begin{align}
    \bar{x}_{wt}^j &= \frac{\sum_i w_{i} x_i^j}{\sum_i w_{i}}, \\
    s_{wt}^j &= \frac{\sum_i w_{i}}{ (\sum_i w_{i})^2 - \sum_i w_{i}^2 } \sum_i w_{i} (x_i^j - \bar{x}_{wt}^j)^2,
\end{align}
where $w_{i}$ denotes the inverse probability weight of subject $i$.

We evaluated covariate balance using the maximum absolute standardized mean difference (MASMD) as follows:
\begin{equation}
    \text{MASMD}(\boldsymbol{d}) = \max_j |d^j|,
\end{equation}
where $\boldsymbol{d}=[d^1,\dots,d^m]^T$.
MASMD is a measure of the bias in the distribution of covariates between the treatment and control groups.
If the bias in the distribution of covariates is the source of bias in the estimate, statistical causal inference methods that obtained smaller MASMDs are likely to have smaller biases in the estimates.
Considering this, when the true treatment effect is unknown, as in Experiment II, MASMD provides a signal of the scale of the bias in the estimate.
For FedCI, MASMD was not available because neither propensity score calculation nor matching was done.

\subsection{Experiment I: Proof-of-concept using artificial data}
\label{sec:exp_I}

As a proof-of-concept of the proposed method, we considered artificial data consisting of 1000 subjects with six covariates:
\begin{equation}
    \boldsymbol{x}_i = [x_i^1, \dots, x_i^6]^T \sim \mathcal{N}(\boldsymbol{0}, S), \quad (i=1,\dots,1000),
    \label{eq:artidata_covariance}
\end{equation}
where $\mathcal{N}(\boldsymbol{0}, S)$ is a normal distribution with mean $\boldsymbol{0}$ and 
covariance matrix $S$ such that $S_{j,j}=1$ and $S_{j,j'}=0.5 ~ (j \neq j')$.
Subject $i$ is treated ($z_i = 1$) with the following probability:
\begin{equation}
    Pr(z_i = 1 | x_i) = \frac{1}{1+\exp \left( \sum_{j=1}^6 -\frac{1}{6}x_i^j \right) }.
    \label{eq:artidata_treatment}
\end{equation}
The outcome $y_i$ is determined as
\begin{equation}
    y_i = \sum_{j=1}^m x_i^j + z_i + \epsilon_i,
    \label{eq:artidata_outcome}
\end{equation}
where $\epsilon_i \sim \mathcal{N}(0,0.1^2)$.
As the treatment increases the outcome by 1, the true ATE is 1 in this artificial dataset.
In this setting each $x_i^j$ is positively correlated with both $z_i$ and $y_i$.
Hence, an estimate of ATE by simply comparing the treatment and control groups is positively biased.
In the data generation process for Experiment I, we first generated $X$ according to (\ref{eq:artidata_covariance}). We then determined $Z$ based on (\ref{eq:artidata_treatment}), and subsequently computed $Y$ according to (\ref{eq:artidata_outcome}).

We focused on ATE in Experiment I.
We considered the benchmark treatment effect $\tau^\text{BM}$ to be 1, the true ATE, and calculated the gap based on it.
We calculated the inconsistency with true based on (\ref{eq:artidata_treatment}).

Dataset $X$ was equally distributed among four parties.
Each party had a dataset $X_{k,l}$ consisting of 500 subjects with three covariates.
In this artificial dataset, the distribution of the covariates was independent of the subject index $i$, and the relationship between the covariates was symmetric.
In this case, the results of the individual analyses of the four parties were virtually equal.
Hence, we considered only the results of the (1,1)th party as a result of the individual analysis.
Furthermore, the same was true for left-side and right-side collaborations, and top-side and bottom-side collaborations.
Therefore, we considered only three types of collaboration: left, top, and whole.
We set $\widetilde{m}_{k,l}=2$, and $\check{m}=3$ for left-side collaboration and $\check{m}=6$ for the top-side and whole collaborations.

\begin{figure}[tb]
  \centering
  \includegraphics[width=\linewidth]{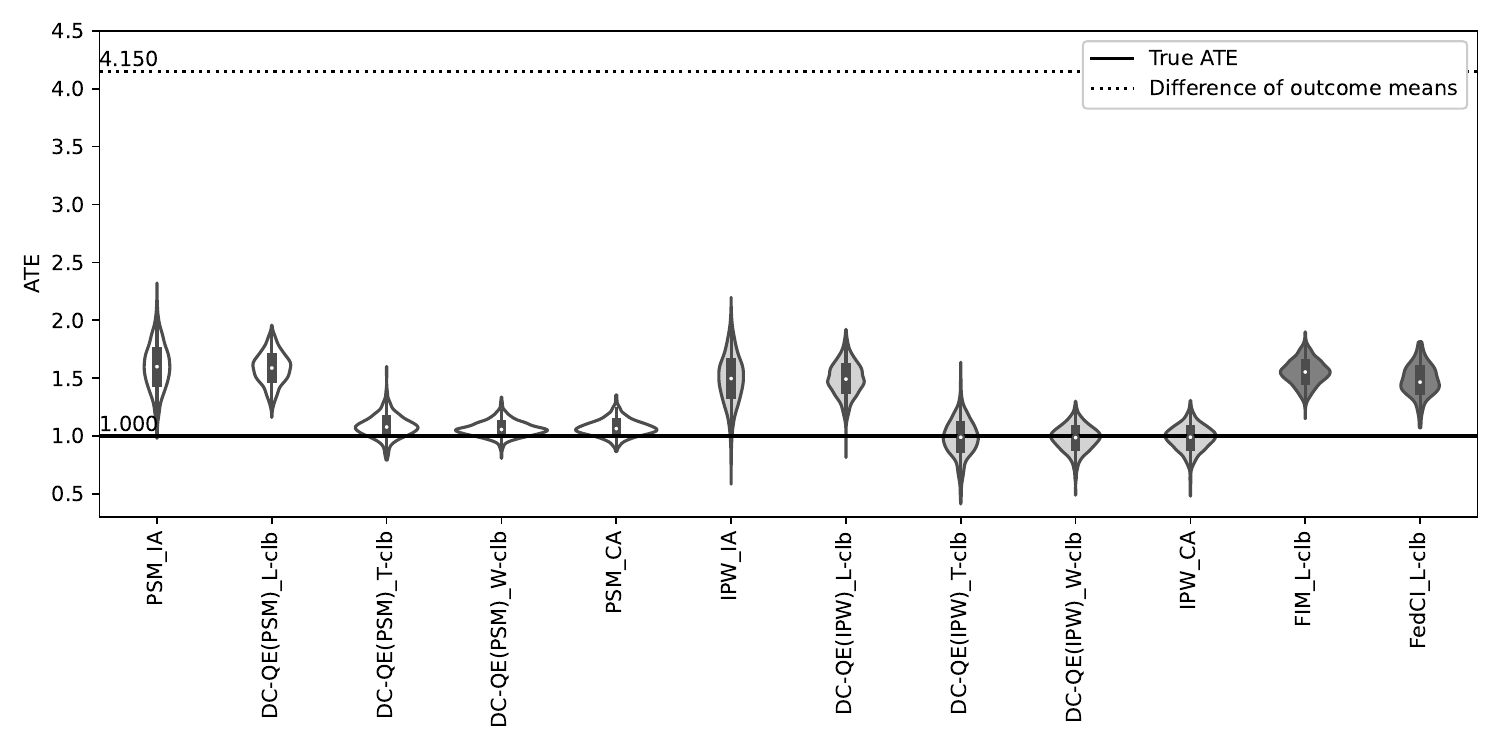}
  \caption{ATE estimates in Experiment I.}
  \label{fig:exp1ate}
\end{figure}

\begin{table}[tb]
\centering
    \scalebox{0.9}{
        \begin{tabular}{llrrrrr}
        \hline\hline
        \multicolumn{1}{c}{\multirow{2}{*}{Estimator}} & \multicolumn{1}{c}{\multirow{2}{*}{Collaboration}} & \multicolumn{1}{c}{\multirow{2}{*}{ATE}} & \multicolumn{1}{c}{\multirow{2}{*}{\begin{tabular}[c]{@{}c@{}}Gap from\\ true ATE (=1)\end{tabular}}} & \multicolumn{2}{c}{Inconsistency} & \multicolumn{1}{c}{\multirow{2}{*}{MASMD}} \\
        \multicolumn{1}{c}{} & \multicolumn{1}{c}{} & \multicolumn{1}{c}{} & \multicolumn{1}{c}{} & \multicolumn{1}{c}{with True} & \multicolumn{1}{c}{with CA} & \multicolumn{1}{c}{} \\ \hline\hline
        PSM & IA & 1.5992 (0.1966) & 0.6306 & 0.0747 (0.0086) & 0.0708 (0.0196) & 0.2714 (0.0826) \\
        DC-QE(PSM) & L-clb & 1.5862 (0.1352) & 0.6016 & 0.0691 (0.0051) & 0.0715 (0.0136) & 0.2508 (0.0577) \\
        DC-QE(PSM) & T-clb & 1.0840 (0.0960) & 0.1275 & 0.0485 (0.0154) & 0.0305 (0.0160) & 0.1097 (0.0404) \\
        DC-QE(PSM) & W-clb & 1.0628 (0.0664) & 0.0914 & 0.0431 (0.0114) & 0.0335 (0.0103) & 0.0790 (0.0290) \\
        PSM & CA & 1.0711 (0.0724) & 0.1015 & 0.0460 (0.0116) & 0.0000 (0.0000) & 0.0744 (0.0285) \\ \hline
        IPW & IA & 1.4928 (0.2094) & 0.5354 & 0.0747 (0.0086) & 0.0708 (0.0196) & 0.2479 (0.0713) \\
        DC-QE(IPW) & L-clb & 1.4943 (0.1432) & 0.5146 & 0.0691 (0.0051) & 0.0715 (0.0136) & 0.2372 (0.0506) \\
        DC-QE(IPW) & T-clb & 0.9843 (0.1556) & 0.1563 & 0.0485 (0.0154) & 0.0305 (0.0160) & 0.0818 (0.0345) \\
        DC-QE(IPW) & W-clb & 0.9805 (0.1091) & 0.1108 & 0.0431 (0.0114) & 0.0335 (0.0103) & 0.0603 (0.0233) \\
        IPW & CA & 0.9820 (0.1088) & 0.1103 & 0.0460 (0.0116) & 0.0000 (0.0000) & 0.0314 (0.0145) \\ \hline
        FIM & L-clb & 1.5507 (0.1081) & 0.5612 & 0.0672 (0.0047) & 0.0809 (0.0144) & 0.2395 (0.0507) \\ \hline
        FedCI & L-clb & 1.4839 (0.1341) & 0.5020 & not available & not available & not available\\
        \hline\hline
        \end{tabular}
    }
\caption{Means and standard errors (in parentheses) of performance measures calculated in Experiment I. ($B=1000$ except FedCI ($B=100$))}
\label{tab:exp1}
\end{table}

First, we demonstrate that the proposed method can reduce biases and random errors of the estimate.
Fig. \ref{fig:exp1ate} shows the estimated ATEs distributions calculated from bootstrap replicates.
The estimated ATEs for PSM and IPW in individual analyses had less bias than the estimated effect ($\approx 4.150$), which was the difference in outcome means between treatment and control groups.
Distributions for top-side collaborations using DC-QE(PSM) and DC-QE(IPW) were closer to the true ATE than for individual analyses using PSM and IPW, respectively.
This implies that biases were reduced by top-side collaborations, which shared intermediate representations constructed from the data where covariates were distributed.
Moreover, the distributions for left-side collaborations using DC-QE(PSM) and DC-QE(IPW) were narrower than those for individual analyses using PSM and IPW, respectively.
This implies that random errors were reduced by left-side collaborations, which shared intermediate representations constructed from data where subjects were distributed.
The distributions for whole collaborations using DC-QE(PSM) and DC-QE(IPW) showed reduction of bias and random errors and seemed comparable to the central analyses.

Results comparison between the proposed method and related methods in Fig. \ref{fig:exp1ate} showed that the left-side collaborations using DC-QE(PSM) or DC-QE(IPW) were comparable to FedCI.
The distribution for FIM was narrower than those.
This implies that random errors for left-side collaborations using DC-QE(PSM) or DC-QE(IPW) were larger than for FIM.
Conversely, the top-side or whole collaboration using DC-QE(PSM) or DC-QE(IPW) seemed to obtain estimates with less bias than FIM and FedCI.

Table \ref{tab:exp1} lists the numerical results of Experiment I for the three aspects.
We quantitatively validated that the proposed method reduced both bias and random errors (shown as standard errors), based on the ATE results, compared with related methods.
Moreover, gaps from the true ATE were sufficiently small to be comparable to those in the central analyses.

Collaborations using DC-QE obtained good estimation results because their propensity scores were estimated accurately.
This can be confirmed by the results of the inconsistencies with true and CA.
Note that inconsistency results were the same regardless of whether the estimation operates with PSM or IPW 
because the propensity scores were estimated before the estimation of treatment effect.
Hence, we discuss inconsistencies only for operations using PSM.
The inconsistencies with true for collaborations using DC-QE(PSM) were smaller than for individual analysis using PSM.
This suggests that more accurate propensity scores were obtained through collaborations.
Moreover, the inconsistency with the CA results showed that the estimated propensity scores in collaborations using DC-QE were close to those in the central analyses.
Compared with FIM, the left-side collaboration using DC-QE was similar to FIM.
Moreover, the top-side or whole collaboration using DC-QE outperformed FIM.

The results of the MASMD for DC-QE show that the covariates were well balanced because the propensity scores were accurately estimated.
The MASMDs in collaborations using DC-QE(PSM) were smaller than in the individual analysis using PSM.
The same was true for the comparison of collaborations using DC-QE(IPW) and individual analysis using IPW.
These results suggest that the estimated propensity scores from the proposed method balanced covariates well.
In the comparison with FIM, the left-side collaboration using DC-QE balanced the covariates to the same degree as FIM.
Moreover, the top-side or whole collaboration using DC-QE balanced better than FIM.

In summary, for artificial data, the proposed method showed better performance than the individual analyses and was comparable to the central analyses.
Conversely, compared with related methods, the left-side collaborations using DC-QE showed comparable results to FIM and FedCI.
Moreover, the top-side or whole collaborations using DC-QE outperformed FIM and FedCI.

\subsection{Experiment II: Evaluation on employment program data}
\label{sec:exp_II}

In Experiment II, we evaluated the proposed method using employment data 
obtained from the National Supported Work (NSW) demonstration project and the Panel Study of Income Dynamics (PSID).
The PSID was a household survey.
The NSW demonstration project was as follows:
``a temporary employment program designed to help disadvantaged workers lacking basic job skills move into the labor market by giving them work experience and counseling in a sheltered environment''\citep{lalonde1986evaluating}.
The task of estimating treatment effects on subjects' incomes using the NSW data has been popular in the causal inference literature since \cite{lalonde1986evaluating}.
In fact, the NSW data are listed in \cite{guo2020survey} as open-source data to facilitate causal inference research.

\cite{dehejia1999causal} estimated the ATT using propensity scores for combined data where the treatment group was composed of treated subjects in the NSW and control group of subjects in PSID.
Subsequently, \cite{dehejia1999causal} compared the estimated ATTs from the combined data with the estimated treatment effects from the NSW data.
Because the NSW data were obtained from a randomized experiment, the estimated treatment effects were benchmarks for the estimated ATTs from the combined data.
\cite{dehejia1999causal} showed that propensity score analysis led to estimates much closer to the benchmarks than those from other methods, such as linear regression.
Following \cite{dehejia1999causal}, we used the NSW data and combined data to evaluate the proposed method.
\footnote{Data was obtained from Rajeev Dehejia's website (\url{https://users.nber.org/~rdehejia/nswdata.html}) on May 1, 2022.}

The NSW data consisted of 185 treatment and 260 control subjects.
The PSID data had 2490 subjects, regarded as controls in the combined data.
Each participant had eight covariates: 
age, education (years of education), married, no degree, Black, Hispanic, re74 (real earnings in 1974), and re75 (real earnings in 1975).
Treatment was 1 if treated and 0 otherwise.
The outcome was re78 (real earnings in 1978).

As in \cite{dehejia1999causal}, we focused on ATT in Experiment II.
We considered an estimated treatment effect ($\approx \$1794$) from the NSW data as the benchmark, which was the difference in outcome means between the treatment and control groups.
As the true ATT was unknown, the gap from the benchmark was only a reference measure.
The inconsistency with true was not calculated because true propensity scores were unknown.

The dataset $X$ was distributed into four parties with 1337 subjects each.
The left-side parties, (1,1)th and (2,1)th, had four covariates: age, married, education, and no degree.
The right-side parties, (1,2)th and (2,2)th, had four covariates: Hispanic, Black, re74, and re75.
The dataset was distributed after the subjects were randomly sorted.
Hence, the distributions of covariates were virtually the same, whether the party was on the top-side or bottom-side.
This implies that the individual analysis result was affected only by whether it was on the left- or right-side.
The same was true for the collaboration results.
Therefore, we considered individual analyses in the (1,1)th and (1,2)th parties, collaboration analyses in left-, right-, top-, and whole collaborations, and central analysis.
We refer to the individual analysis in (1,1)th as the left-side individual analysis, and in (1,2)th as the right-side individual analysis.
We set $\widetilde{m}_{k,l}=3$, and moreover $\check{m}=4$ for the left- and right-side collaborations, and furthermore $\check{m}=8$ for the top- and whole collaborations.

\begin{figure}[tb]
  \centering
  \includegraphics[width=\linewidth]{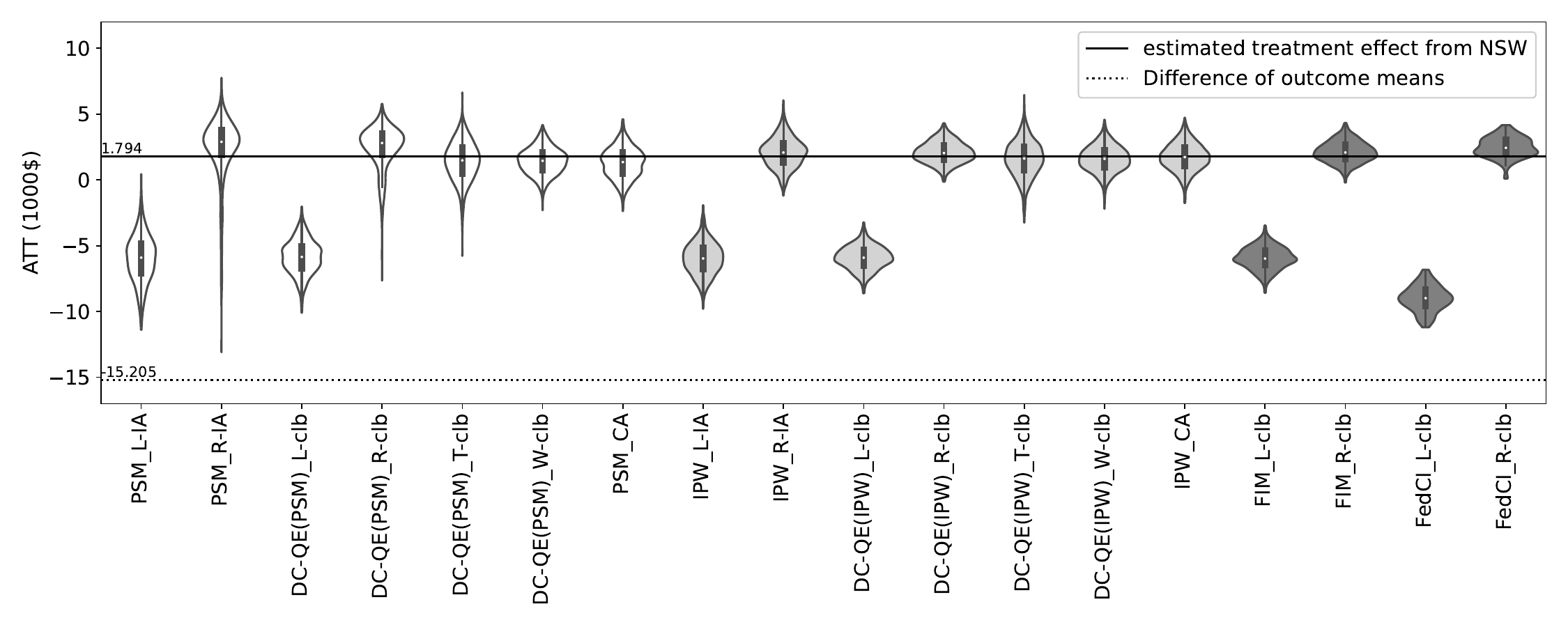}
  \caption{ATT estimates in Experiment II.}
  \label{fig:exp2att}
\end{figure}

\begin{table}[tb]
\centering
    \scalebox{0.9}{
        \begin{tabular}{llrrrr}
        \hline\hline
        \multicolumn{1}{c}{Estimator} & \multicolumn{1}{c}{Collaboration} & \multicolumn{1}{c}{ATT (1000\$)} & \multicolumn{1}{c}{\begin{tabular}[c]{@{}c@{}}Gap from\\ ATT in NSW ($\approx$1.794)\end{tabular}} & \multicolumn{1}{c}{Inconsistency   with CA} & \multicolumn{1}{c}{MASMD} \\ \hline\hline
        PSM & L-IA & -5.9910 (1.7841) & 7.9870 & 0.1281 (0.0120) & 1.4280 (0.2394) \\
        PSM & R-IA & 2.4057 (2.3567) & 2.4336 & 0.0970 (0.0109) & 1.6169 (0.5565) \\
        DC-QE(PSM) & L-clb & -5.8610 (1.2628) & 7.7587 & 0.1274 (0.0085) & 1.3713 (0.1573) \\
        DC-QE(PSM) & R-clb & 2.4524 (1.6289) & 1.7560 & 0.0967 (0.0076) & 1.4605 (0.4805) \\
        DC-QE(PSM) & T-clb & 1.4273 (1.4604) & 1.5051 & 0.0231 (0.0096) & 0.3671 (0.1306) \\
        DC-QE(PSM) & W-clb & 1.4196 (0.9870) & 1.0553 & 0.0240 (0.0068) & 0.3022 (0.0974) \\
        PSM & CA & 1.3000 (1.1258) & 1.2290 & 0.0000 (0.0000) & 0.2842 (0.0871) \\ \hline
        IPW & L-IA & -5.9773 (1.2099) & 7.8652 & 0.1281 (0.0120) & 1.3694 (0.1385) \\
        IPW & R-IA & 2.0710 (1.1110) & 1.1444 & 0.0970 (0.0109) & 1.2722 (0.2020) \\
        DC-QE(IPW) & L-clb & -5.9188 (0.8732) & 7.7624 & 0.1274 (0.0085) & 1.3603 (0.1009) \\
        DC-QE(IPW) & R-clb & 2.0780 (0.7972) & 0.8458 & 0.0967 (0.0076) & 1.2301 (0.1372) \\
        DC-QE(IPW) & T-clb & 1.5931 (1.3337) & 1.3481 & 0.0231 (0.0096) & 0.2844 (0.1084) \\
        DC-QE(IPW) & W-clb & 1.6078 (0.9663) & 0.9837 & 0.0240 (0.0068) & 0.2362 (0.0771) \\
        IPW & CA & 1.7320 (1.0190) & 1.0204 & 0.0000 (0.0000) & 0.1777 (0.0566) \\ \hline
        FIM & L-clb & -5.9516 (0.8424) & 7.7916 & 0.1271 (0.0085) & 1.3641 (0.0975) \\
        FIM & R-clb & 2.1236 (0.7873) & 0.8530 & 0.0963 (0.0077) & 1.2383 (0.1374) \\ \hline
        FedCI & L-clb & -9.0077 (0.9576) & 10.8440 & not available & not available \\
        FedCI & R-clb & 2.4723 (0.7729) & 1.0252 & not available & not available\\
        \hline\hline
        \end{tabular}
    }
\caption{Mean and standard errors (in parentheses) of performance measures calculated in Experiment II. ($B=1000$ except FedCI ($B=100$))}
\label{tab:exp2}
\end{table}

Fig. \ref{fig:exp2att} shows the distributions of the estimated ATTs calculated from the bootstrap replicates.
As with the results of Experiment I, whole collaborations using DC-QE(PSM) and DC-QE(IPW) seemed to obtain the estimated ATT close to the central analyses.
Compared with related methods, the left- and right-side collaborations using DE-QE(IPW) obtained estimates comparable to FIM and FedCI, but using DE-QE(PSM) exhibited worse results.

Table \ref{tab:exp2} lists the numerical results of Experiment II.
Since the true ATT was unknown, we did not check whether the biases were smaller in the collaborations than in the individual analyses.
Conversely, we confirmed that random errors were reduced for left- and right-side collaborations using DC-QE compared to left- and right-side individual analyses, respectively.
The gap results showed that, in operations with PSM, the collaborations were close to the benchmark estimate.
However, in operations with IPW, the gap in the top-side collaboration was larger than in the right-side individual analysis.
Note that as the benchmark was not the true ATT, the gap increase due to collaboration did not necessarily represent a deterioration in performance.
Compared with related methods, as mentioned above, the left- and right-side collaborations using DE-QE(IPW) obtained estimates comparable to FIM and FedCI.

In view of this, we should also evaluate the inconsistencies and MASMD performances.
The inconsistency results showed that the estimated propensity scores from collaborations using DC-QE were closer to those from the central analyses than to the individual analyses.
The MASMD results showed that the estimating from collaborations using DC-QE led to more balanced covariates than those from individual analyses.
In particular, MASMDs in the top-side collaborations using DC-QE were smaller than in individual analyses.
Hence, the top-side collaborations using DC-QE led to estimates with less bias, although the benchmark gaps were large to some extent.
The same is true for the results in whole collaborations using DC-QE.
Compared with related methods, the left- and right-side collaborations using DE-QE obtained inconsistencies and MASMDs comparable to FIM and MASMD.
Moreover, the top-side and whole collaborations using DE-QE obtained better inconsistencies and MASMDs than FIM.

In summary, for real-world data, the proposed method showed better performance than the individual analysis and was comparable to the central analysis.
Compared with related methods, the left- and right-side collaborations using DE-QE obtained estimates comparable to FIM and FedCI.
Moreover, the top-side and whole collaborations using DC-QE outperformed FIM in inconsistencies and MASMD.


\section{Discussion}
\label{sec:discussion}

In this section, we discuss our results, focusing on some of the validity threats discussed by \cite{feldt2010validity}: credibility, internal validity, construct validity, and external validity.

We first discuss credibility, which focuses on the reasons we are confident that our results are correct.
In preparing the numerical experiments, we removed the code bugs by assigning different authors to write and review the code.
This implies that our results have high credibility.

Second, we discuss internal validity, which focuses on how sure we can be that the proposed method improved the performance.
To achieve valid results, we simplified the experimental design to reduce the influence of factors other than those of interest.
The setting $c=2$ and $d=2$ was the minimum situation to study the performance of the proposed method to horizontal and vertical partitioned data.
Moreover, we conducted many trials with the bootstrap method in the two experiments to obtain robust results.
This led to robust results that accounted for perturbations in data distribution.
Then, we obtained consistent results throughout the two experiments that the proposed method outperformed the individual analysis.

Third, we discuss construct validity, which focuses on the relation between the theory behind the two experiments and our results.
As mentioned in Section \ref{sec:relatedwork}, we expected that estimating with more subjects and covariates would lead to reduced random errors and biases, respectively.
We observed in both experiments that the proposed method resolved the lack of covariates and resulted in significant improvements in performance over individual analyses.
Conversely, the proposed method resolved the lack of subjects and improved the performance over individual analyses, but not as much as when the lack of covariates was resolved.
Moreover, performances with FIM and FedCI were also not as good as when the proposed method resolved the lack of covariates.
Hence, if collaboration for data where covariates are distributed is possible, the proposed method will produce more reliable results than the other methods.
As mentioned in Section \ref{sec:method_ad_disad}, dimensionality reduction caused performance degradation in the two experiments.
This was confirmed by the fact that collaboration performances by all parties using the proposed method were lower than in centralized analysis.
Conversely, collaboration performances using the proposed method were better than in individual analyses.
This implies that the proposed method resolved the lack of subjects and covariates and then improved performance sufficiently to overcome the degradation caused by dimensionality reduction.

Fourth, we discuss external validity, which focuses on whether we can generalize the results outside our study.
We cannot guarantee the robustness of our conclusions to other datasets and distributional settings because we conducted the experiments only for two datasets and simplified the experimental design.
This is the main limitation of this study.
To examine the effect of dataset type, for example, medical or industrial, the list suggested by \cite{guo2020survey} may be useful.
\footnote{The list can be accessed at \url{https://github.com/rguo12/awesome-causality-data}.}
For the distribution setting, we can consider, for example, cases where subjects are not independent of each other and where the parties have different distributions of data.

\section{Conclusion}
\label{sec:conclusion}

In this study, we proposed DC-QE, which enables causal inference from distributed data with privacy preservation.
The proposed method preserves privacy of private data by sharing only dimensionality-reduced intermediate representations.
Moreover, the proposed method can reduce both random errors and biases, whereas existing methods can only reduce random errors in estimating treatment effects.
The proposed method exhibited good performance in our experiments.

The proposed method is a highly flexible framework with applications in terms of data-sharing and estimation.
Intermediate representations can be published for researchers instead of private data.
In the estimation, researchers can select methods, such as DC-QE(PSM) and DC-QE(IPW), according to their needs.
If the proposed method is widely known and used, a platform for publishing intermediate representations generated in studies as open data may be required.
Many researchers can then use these published intermediate representations to find causalities and accumulate a knowledge base.
Notably, the proposed method is an application of machine learning methods to statistical causal inference; hence, various unsupervised or supervised learning methods can be used to construct intermediate or collaborative representations.

In future studies, we will investigate performance in cases where treatments and outcomes are categorical.
Another important direction for future study would be to extend the proposed method to address causal graphs.


\section*{Acknowledgement}

The authors would like to thank the anonymous reviewers for their constructive comments.
This work was supported in part by the New Energy and Industrial Technology Development Organization (NEDO), (No. JPNP18010), Japan Science and Technology Agency (JST), (No. JPMJPF2017), Japan Society for the Promotion of Science (JSPS), and Grants-in-Aid for Scientific Research (Nos. JP21H03451, JP22K19767, JP22H00895, JP23H00462, JP23H03411).
We would like to thank Editage (\url{www.editage.com}) for English language editing.

\bibliography{ref}

\end{document}